\documentclass[twocolumn,journal]{IEEEtran}
\usepackage[T1]{fontenc}
\usepackage[utf8]{inputenc} 
\usepackage[unicode=true,
 bookmarks=true,bookmarksnumbered=true,bookmarksopen=true,bookmarksopenlevel=1,
 breaklinks=false,pdfborder={0 0 0},pdfborderstyle={},backref=false,colorlinks=false]
 {hyperref}
\hypersetup{pdftitle={Your Title},
 pdfauthor={Your Name},
 pdfpagelayout=OneColumn, pdfnewwindow=true, pdfstartview=XYZ, plainpages=false, breaklinks=true}
 \usepackage{url}

\usepackage{breakurl}

\makeatletter
\let\oldforeign@language\foreign@language
\DeclareRobustCommand{\foreign@language}[1]{%
  \lowercase{\oldforeign@language{#1}}}

\usepackage[caption=false,font=footnotesize]{subfig}

\makeatother

\begin{document}
\title{Tracing Contacts\\to Control the COVID-19 Pandemic}
\author{Christoph, Michael and Daniel Günther\thanks{Christoph Günther is with the German Aerospace Center, 82234 Weßling,
and with Technische Universität München, 80330 Munich, Germany, e-mail:
\protect\href{http://KN-COVID\%40dlr.de}{KN-COVID@dlr.de}.}\thanks{Michael and Daniel Günther are students at Technische Universität
München, 80330 Munich, Germany, e-mail: \protect\href{http://m.guenther\%40tum.de\%2C\%20d.guenther\%40tum.de}{m.guenther@tum.de, d.guenther@tum.de}.}}
\markboth{}{}
\IEEEpubid{}
\maketitle
\begin{abstract}
The control of the COVID-19 pandemic requires a considerable reduction
of contacts mostly achieved by imposing movement control up to the
level of enforced quarantine. This has lead to a collapse of substantial
parts of the economy. Carriers of the disease are infectious roughly
3 days after exposure to the virus. First symptoms occur later or
not at all. As a consequence tracing the contacts of people identified
as carriers is essential for controlling the pandemic. This tracing
must work everywhere, in particular indoors, where people are closest
to each other. Furthermore, it should respect people's privacy. The
present paper presents a method to enable a thorough traceability
with very little risk on privacy. In our opinion, the latter capabilities
are necessary to control the pandemic during a future relaunch of
our economy.
\end{abstract}


\IEEEpeerreviewmaketitle{}

\section{Introduction}

\IEEEPARstart{T}{he} COVID-19 pandemic has spread all over the world.
It has already lead to a very large number of fatalities, more than
40'000 as of end of March 2020. The first priority of humanity is
to take all possible actions to prevent more people from dying. In
some places, this lead to enforcing a quarantine on large portions
of the population. The economic damage is substantial. The US alone
is investing USD 2'000 Billions to alleviate the consequences of the
pandemic. Thus, limiting the economical damage by restarting the economy
as soon as possible, while at the same time protecting people, is
of immense importance. The present document aims at contributing specific
suggestions on how to achieve this.

Three important properties of the COVID-19 pandemic are that
\begin{itemize}
\item The sickness is limited to roughly three weeks in time. After this
period, people are either healthy again, hopefully without impairments,
or dead. All evidence expressed publicly, so far, indicates that former
carriers of the disease are not contagious anymore after that time.
A strictly observed quarantine of three weeks has thus the potential
to basically eliminate all carriers of the disease. A quarantine is
never perfect, e.g. due to the need to restock food supplies. As a
consequence some chains of infection will persist.
\item The spreading of the disease in the population is characterized by
an exponential growth. The characteristic parameter $R_{0}$, which
describes the number of people infected by a single carrier, is estimated
to be around 2-2.3. Any value above 1 leads to a exponential growth,
as long as there is no substantial immunity. More detailed epidemiological
models are more differentiated but show a similar threshold behavior\cite{HeiBuc20}.
The value of $R_{0}$, mentioned above is determined by the period
during which a carrier is contagious, the probability of transmitting
the disease, and the number of contacts that the carrier had during
that time. There is no means to control the first factor. The second
may be somewhat influenced by carrying masks but not to a level considered
sufficient. Thus, the most important option for controlling $R_{0}$
is to reduce the contacts between carriers and other citizens.
\item The diagnoses of sick people is a critical element. Some people do
not show symptoms that they associate with the sickness but are nevertheless
infectious. They may be a cause for requiring a longer quarantine
than described above. In addition and most importantly, no one shows
symptoms before being infectious, which means that as long as there
are no tests that everyone can apply at regular intervals, there will
always be a delay before the spreading by a particular individual
can be discontinued. Furthermore, extensive testing as practiced and
further expanded in Germany will be most effective if the most likely
carriers are being tested.
\end{itemize}
Currently, there is a variety of attempts to contain the pandemic,
which should all be followed in parallel. The development of vaccines
and of medications are essential but may not be available in the near
future. This has led to an enforced reduction of contacts by various
levels of quarantine. The concept of achieving immunity by letting
the epidemic spread have rightfully been abandoned, due to the heavy
toll in human lives. Bill Gates formulates what most of us think ``But
bringing the economy back ... that\textquoteright s more of a reversible
thing than bringing people back to life. So we\textquoteright re going
to take the pain in the economic dimension \textemdash{} huge pain
\textemdash{} in order to minimize the pain in the diseases-and-death
dimension.''\footnote{https://www.cnbc.com/2020/03/25/what-bill-gates-would-do-to-fight-covid-19-if-he-were-us-president.html}

The ``how'' of restarting the economy remains. Some authors studied
the effect of relaxing the quarantine at the cost of a regrowth of
infected people before shutting down again \cite{FerLayNedIma20}.
This leads to an increasing level of immunity in a series of waves.
In view of the small percentage of people that are immunized at each
step and in view of the risk of an unmanageable growth, the number
of waves needs to be substantial. Furthermore, each wave costs lives.
China, South Korea and to a much smaller scale Webasto in Germany
have shown an alternative, which consists in a careful tracing of
contacts, associated with testing, and quarantining positively tested
people. We will call these people ``carriers'' throughout the paper.

\section{Tracing Contacts}

Tracing contacts is a rather natural concept for containing the pandemic.
It aims at identifying and subsequently isolating people, who might
potentially be carriers. Since the incubation time until an infected
person becomes infectious herself is around 3 days and since first
symptoms only occur after 5 days at the earliest, with a diagnosis
available at en even later time, there is a lag during which infectious
carriers continue spreading the disease. Thus, knowing contacts to
people who have been identified as carriers, allows isolating unidentified
potential carriers. The frequent absence of clear symptoms is a second
critical cause for the spreading of the infection. In this case contact
tracing allows identifying carriers without symptoms through their
contact to people with symptoms. In that case, the carrier with symptoms
is not the originator but rather helps discovering the originator.
Independent on who is the originator, contact tracing and subsequent
isolation eliminates sources of disease spreading. An immediate testing
and determination of contacts allows to identify further contact whenever
the outcome of the test was positive. In the case of a negative outcome,
testing is repeated after an incubation time, with isolation being
lifted in the case of a second negative outcome.

At my institute (160 people), we traced a number of contacts and noticed
that the complexity of a manual process becomes quickly unmanageable.
Due to the exponential character of the network of relations, there
are simply too many contacts to be traced. We ended up isolating everyone
first at the Institute level, shortly after that and independently
of us at DLR level and finally at national level. This observed complexity
led us to the conclusion that automatic means of tracing are essential.
Raskar et al. \cite{RasSchBarVilc20} have analyzed an approach based
on locating people with a particular focus on privacy-protection and
self-protection against the disease. We follow a somewhat different
approach. It is primarily based on contacts, rather than on locations,
although locations may be used in addition. Furthermore, it is focusing
on the control of the pandemic as a whole. The protection of the individual
turns out to reach a similar level as in the approach by Raskar et
al. \cite{RasSchBarVilc20}.

The present exposition is developed against the background of German
regulations. The public authorities responsible for health is the
``Gesundheitsamt.'' The Gesundheitsämter (many of them, distributed
all over the country) register every person affected by the pandemic
and organize the testing of people. Thus the identity of any person
which either has symptoms, is tested positively or is affected by
the disease is currently known to the local Gesundheitsamt. We shall
subsequently just speak of the Gesundheitsamt as if it was a single
entity. That Gesundheitsamt is a trusted authority independent of
any use of electronic means to trace contacts. It shall thus also
be the trusted authority in our approach, which will be responsible
to operate the server needed to manage the list of carriers. They
do furthermore manage people in quarantine, who have to follow strict
rules in Germany. Not doing so may lead to fines and imprisonment
\cite{Quarantine20}. Additionally, Germany has imposed limitations
on the movement of people, which should not be confused with the stricter
quarantine. In our view, it should be acceptable that regaining new
degrees of freedom may be associated with certain restrictions, which
ensure the traceability of contacts, without unduly exposing privacy.
Recent polls in Germany show a high level of acceptance of restrictions
to combat the pandemic. It may well be acceptable to enforce the use
of tracing, although this is not the focus of the paper.

\section{Technical Implementation}

The precondition for traceability is to use of a smartphone running
a COVID-19 tracing app (the app) or alternatively the use of a low
cost device. For simplicity, the focus of the exposition will be on
an app running on a smartphone. Every person leaving their home shall
be requested to carry such a device, with the app installed and active.
This might be an expectation, which people are free to follow or not.
Whatever solution is preferred is a political decision. The main elements
of its implementation are
\begin{itemize}
\item The automated creation of a list of contact instances my\_ctc, maintained
in the personal device of the user. The number of such entries could
be up to a few thousand entries per day as soon as big events take
place again.
\item The maintenance of a list of infectious carrier of the disease ga\_icd
on the server of the Gesundheitsamt, currently around 70'000 entries
in total with a growth rate of less than 4000 per day.
\item The search for entries from the personal list my\_ctc in the list
ga\_icd retrieved from the Gesundheitsamt.
\item In the case of a hit, the app informs the server of the Gesundheitsamt
about the identifier found.
\item The server and the app cooperate in classifying the category of the
contact (Category 1 or 2, see below). The associated contact persons
might be involved in this classification process.
\item Based on the result, the Gesundheitsamt decides about the quarantining
and testing of the device's owner.
\end{itemize}
The best possible cooperation of the contacts and the Gesundheitsamt
in assessing the category of the contact reduces both the test load
and the necessity of a quarantine. In an initial phase, this may include
the indication of the seat used on a joint train ride, the confirmation
of a joint lunch or the like. Clearly, further technical developments
in sensing of both the mutual placement and orientation of people
will be of great help in automating this process but are not needed
in an initial phase. Such developments could follow similar lines
as the work for indoor position, which achieves high levels of accuracy
\cite{MoaLiShi19}.

\subsection{Actors}

The above description identified a number of actors. Before entering
into this discussion, it is useful to differentiate three categories
of contacts \cite{Category20}:
\begin{itemize}
\item Category 1 contacts are those to which a face-to-face contact accumulated
to more than 15 minutes.
\item Category 2 contacts are those to which a face-to-face contact accumulated
to less than 15 minutes.
\item Uncritical contacts are all others.
\end{itemize}
The consequences of being a Category 1 or 2 contact are defined by
the Gesundheitsamt and may be changed over time. Both categories are
quarantined. Currently, the main difference is in the level of testing.
The Category 2 defines the sampling time of our contact monitoring.

With this preparation,we have the following actors:
\begin{itemize}
\item The Gesundheitsamt (trusted authority): it tests people for COVID-19
infection, it publishes an anonymized list of carriers and it facilitates
the categorization of contacts.
\item Roaming users: their devices monitor contacts at regular intervals
(30 second) and store the list of contacts my\_ctc as well a a list
with location and orientation information my\_loc, their devices check
whether there was a contact to an infected person (at least once per
day), and provide support to the categorization of the contacts, potentially
using location and orientation information. Note that all information
is kept locally with the exception of information exchanged in the
categorization of a contact.
\item Users tested positively: their devices provide their lists my\_ctc
as far back as their owner's infection may have been contagious to
the Gesundheitsamt, they go into treatment or at least quarantine,
and cooperate in determining the category of contacts that they had.
The device uses the list my\_loc to support the classification of
contacts to other people. Although the position information is kept
locally, it is partially disclosed to the Gesundheitsamt in the assessment
of contacts' categories.
\item Users with a critical contact (Category 1 or 2): they also go into
quarantine and are subject to an immediate test. In the case of a
positive outcome, they change category. Otherwise, they are tested
again after an incubation time. In a second negative testing, they
are freed from quarantine obligations.
\end{itemize}

\subsection{Tracing Method}

There is a number of options to detect the proximity of people. We
propose to use Bluetooth transceivers to send beacons and monitor
for such beacons at regular intervals. The benefit of using Bluetooth
is that corresponding interfaces are included in nearly every smartphone
and that they are furthermore available on cheap platforms. In addition,
Bluetooth creates a direct relationship between the potential contact
persons, which works everywhere, including shopping malls or the underground
metro station. Although not too reliable, the power level can be used
as an indication of the distance between the transmitter and the receiver,
and could thus be used as a filter. The details of this aspect need
further assessment. Furthermore, the use of Bluetooth is associated
with a low power consumption. The proposal made in Section \ref{sec:Testing}
uses functions available in the Application Programming Interfaces
(API) of Android and Apple iOS. More refined solutions may be implemented
by Google and Apple, themselves providing improved power management,
relative contact positioning, safety against manipulation and the
like.

Tracing may either be performed on a voluntary basis or enforced.
The knowledge of being a carrier (positive testing) does not provide
benefits to people without or with marginal symptoms. It rather puts
them into quarantine and thus reduces their freedom of movement. Quarantining
carriers has a huge benefit for society, however. Thus, the incentives
to individuals are purely ethical, which seems to be sufficient at
the time. Thus, we focus on the voluntary approach but provide some
hints for enforcement as well.
\begin{itemize}
\item In a preparatory phase, the user installs the tracing app. In the
case of enforcement, the app creates a connection of data from an
official ID-card and the device and then registers the user with that
data. This creates a permission to roam and is communicated to the
mobile operator. It can furthermore be used to prevent a number of
manipulations to evade quarantining, for example. In the case of a
voluntary roll out, this registration does not exist, and even in
the mandatory case, it is only used to prevent manipulations and does
in particular not create any additional means of tracking.
\item Every day, the app chooses a random daily identifier my\_rdi, which
it broadcasts at regular intervals using a Bluetooth protocol (see
Section \ref{sec:Testing}). The identifier provided by the device
is C0F1D19|my\_dri. The randomness of the my\_rdi prevents any correspondence
with a particular device or user. It is changed daily to prevent tracking
by any fixed monitoring stations.
\item In parallel, the device searches for the beacons of other devices.
This monitoring is performed every 30 seconds. Whenever the device
detects an identifier of the form C0F1D19|fg\_dri for the first time,
it adds fg\_dri to its list my\_ctc and stores the current time (in
30 second units). If it sees the identifier again, it updates the
duration of the contact. In total, there are 720 two-minutes intervals
in 24 hours. Assuming that someone is surrounded by up to 6 people
during 12 hours would lead us to 2160 entries. There is no difficulty
in storing that number, but this exposes the importance of applying
simple filtering to control the complexity of later processing steps.
\item Whenever the Gesundheitsamt updates its list ga\_ctc, which is signed
using its private key, the device checks for matches between ga\_ctc
and my\_ctc. The increase in carriers is around 4000 per day in Germany.
The list shall include these entries as well as those of the day before,
which is perfectly manageable. The random device identifier and the
date must both match, since the identifier is changed every day. Note
that a very high level of anonymity is preserved up to this point.
\item If there are matches in the device's list my\_ctc and in the list
of the Gesundheitsamt ga\_ctc, there are two different options:
\begin{itemize}
\item The devices notifies its owner and asks him about his preferences.
If the preference is to enter quarantine without further checking,
no further action is needed and no information is ever exchanged.
\item In all other cases, the Gesundheitsamt and the device aim at categorizing
the contact. This requires a negotiation, which can be handled by
a mailbox to prevent the disclosure of the person's identity. In advanced
negotiations, the information from my\_loc will typically be used
in the process of categorization.
\end{itemize}
\item Once the category of a contact is determined, the Gesundheitsamt either
asks the person to quarantines herself and organizes testing, or just
drops the alert if the contact was uncritical. In the latter case,
no further data is exchanged and the data associated with the inquiry
is erased.
\item In the case of a critical contact, the Gesundheitsamt invites the
person for testing. All exchanges can again be handled through a mailbox.
This does again not require the disclosure of the identity of the
person. If the testing is twice negative, the person leaves the quarantine
and the data is erased.
\item In the case of a positive testing, the app provides the contact history
my\_ctc from the beginning of the estimated infection period to the
Gesundheitsamt. The disclosure of the identity of the person is not
needed for pandemic control. The app maintains the list of locations
from the estimated infection onward in order to respond to further
inquiries from the Gesundheitsamt.
\item The device continues comparing its list my\_ctc with later provisions
of ga\_ctc. This is necessary, due to the significant delay before
some carriers are found and since it is the last contact, which is
determining the end of the quarantine period.
\item Whenever the Gesundheitsamt receives a list of my\_ctc including the
timing and the duration, it will add the random identifiers to its
list ga\_ctc. Depending on the evolution of the pandemic and future
experience, it may decide to only trace contacts to Category 1 or
to both categories. It will add these contacts to its list and publish
a signed copy of ga\_ctc at regular intervals. As a consequence, listed
identifiers will trigger an inquiry of the associated devices with
the Gesundheitsamt to ask for categorization. Once every user device
has performed its matching, there will be no unidentified hits in
the past. Thus, the Gesundheitsamt can erase all non-public information
associated with the published list. Since some devices may not have
contact to the Gesundheitsamt for a few hours, there should be a margin
in erasing this data, e.g. one extra day.
\end{itemize}

\subsection{Tracking}

From an epidemiological perspective, users that are quarantined would
ideally be tracked. The procedure is straight forward: whoever leaves
the location of the quarantine is warned. In the case of a continued
breach of rules, the Gesundheitsamt is informed and takes action.
From this time onward, the person could be continuously tracked to
support her repatriation into her quarantine zone. This is certainly
controversial and not too compatible with a voluntary tracing. It
may be activated if enforcement of tracing turns out to be necessary.
Currently, this seems not to be the case.

\section{Threats}

The tracing described above is meant to control the pandemic and to
enable a restart of the economy, while keeping citizens as protected
as possible. In the case of a voluntary use of the system, the main
threats are attacks on the privacy of users. They are not only serious
but may additionally jeopardize the acceptance of tracing as a method
to control the pandemic. In the case of enforced tracing, there are
additionally options for evading tracing or tracking. This is mentioned
but not discussed in any depth.

\subsection{Attacking Privacy and other Misuses}

The primary line of attack to access the personal profile of a particular
person is through the app. Thus, the app needs a thoughtful design
and implementation. This is, however, a requirement, which it shares
with any other software using personal data and localization. A similar
statement holds for the software run on the server of the Gesundheitsamt.
It should avoid any deficiencies but is still exposed to exploits
of the operating systems and the like. We also assume that the public
key cryptosystem is secure in the relevant time. The data base of
the Gesundheitsamt is only of limited interest, since it contains
very little information and since the data is not personalized. The
bigger threat is the impersonation of the Gesundheitsamt, it may lead
to a number of options, which mostly don't have a clear benefit, like
\begin{itemize}
\item The removal or addition of contacts.
\item The false categorization of contacts.
\item The undue convocation of people to testing.
\item The quarantining of healthy people.
\end{itemize}
The most influential possibility is to add a carrier to ga\_ctc and
to thus retrieve the list of his contacts. This, however, requires
finding a valid random daily identifier, e.g. by creating an explicit
contact to a person as well as a major software bug at the Gesundheitsamt
e.g. by exposing its private key. Other sophisticated attacks are
conceivable, e.g. using a network of cooperating Bluetooth units to
profile users by tracking their passage near those units. This is
not particular to the present system, however. Otherwise, we did not
find an obvious other critical attack so far. In the end the usefulness
of tracing carriers of COVID-19 and of restoring normality to our
daily life have to be balanced against fears of potential attacks.

\subsection{Escaping Control}

The consequence of having been in contact with an infected fellow
citizen is to become quarantined. Some people may want to avoid that,
even in the case of enforcement. Most options such as roaming without
an active device, breaking quarantine rules, using different devices,
uninstalling and reinstalling the app or cheating during the categorization
can all be handled by appropriate measures. They will have to be addressed
if enforcement is really desired. This is currently not the case.

\section{Prototype Implementation\label{sec:Testing}}

An implementation of the above system could easily be performed by
the companies Google for Android devices and Apple for iOS devices.
A more detailed design will need a further specifications of the protocols,
which should be done jointly to achieve the fastest possible availability
of a fully inter-operable system. We studied different mechanisms
provided by Bluetooth in Application Programming Interfaces (API).
iBeacons, which is a protocol used for indoor location services, became
our initial candidate. This protocol allows devices to broadcast identifiers,
which are received by other devices in the neighborhood. The received
signal strength can be used as an indicator of the transmitter to
receiver distance. The concentration of transmission and monitoring
around 30 seconds intervals of the time of the day can be used to
implement a simple form of power management.

The focus of our testing was on verifying the possibility of using
a mechanism provided by an API. Thus, we implemented an app on iOS
to transmit iBeacons and used the nRF Connect for Mobile app to monitor
these beacons. This worked whenever the app was in the foreground
of the iOS device. The transmission was, however, discontinued, whenever
the app was sent into the background. As a consequence, we implemented
an alternative approach using the standard Bluetooth Low Energy (BLE)
protocol. A corresponding app was written for iOS and and another
one for Android. Both apps implement the beacon transmission and beacon
monitoring. The source code can be downloaded from https://github.com/danielgnt.
The subdirectories bletrack-android and bletrack-ios contain the associated
code. These apps could successfully monitor beacons between Android
phones as well as between iOS and Android phones. All associated trials
worked with the apps in the background on both phones. However, we
could not get the iOS to iOS scenario working with both apps in the
background. It only works when one of the apps is in the foreground,
which is not sufficient. If this could be solved, a large community
of programmers could implement the tracing system described above.

\section{Conclusions}

The present paper exposes an automated, privacy preserving, tracing
method based on Bluetooth radio contacts, which consequently works
indoors, where people come closest to each other. The approach uses
random daily identifiers to trace contacts. The randomness and daily
updates prevent most attacks on privacy. The information needed to
trace contacts is maintained locally in the personal device. The health
agency ``Gesundheitsamt'' is a trusted authority, which only stores
contact profiles of positively tested people. This data does not have
to include any means of identification of physical person.

The next step in bringing this approach to reality would be to setup
a task force force designing the details of the protocol, as well
as implementing and testing the mobile and server components. The
aim should be for a quick and stable initial operational systems.
The outcome should be further optimized in a second phase to improve
contact classification in order to reduce unnecessary testing and
quarantining.

\bibliographystyle{IEEEtran}
\bibliography{Contact_Tracing_A}

\begin{thebibliography}{1}
\providecommand{\url}[1]{#1}
\csname url@samestyle\endcsname
\providecommand{\newblock}{\relax}
\providecommand{\bibinfo}[2]{#2}
\providecommand{\BIBentrySTDinterwordspacing}{\spaceskip=0pt\relax}
\providecommand{\BIBentryALTinterwordstretchfactor}{4}
\providecommand{\BIBentryALTinterwordspacing}{\spaceskip=\fontdimen2\font plus
\BIBentryALTinterwordstretchfactor\fontdimen3\font minus
  \fontdimen4\font\relax}
\providecommand{\BIBforeignlanguage}[2]{{%
\expandafter\ifx\csname l@#1\endcsname\relax
\typeout{** WARNING: IEEEtran.bst: No hyphenation pattern has been}%
\typeout{** loaded for the language `#1'. Using the pattern for}%
\typeout{** the default language instead.}%
\else
\language=\csname l@#1\endcsname
\fi
#2}}
\providecommand{\BIBdecl}{\relax}
\BIBdecl

\bibitem{HeiBuc20}
M.~an~der Heiden and U.~Buchholz, ``{Modellierung von Beispielszenarien der
  SARS-CoV-2-Ausbreitung und Schwere in Deutschland},'' 2020.

\bibitem{FerLayNedIma20}
N.~Ferguson, D.~Laydon, G.~Nedjati-Gilani, N.~Imai, K.~Ainslie, M.~Baguelin,
  S.~Bhatia, A.~Boonyasiri, Z.~Cucunuba, G.~Cuomo-Dannenburg \emph{et~al.},
  ``{Impact of non-pharmaceutical interventions (NPIs) to reduce COVID-19
  mortality and healthcare demand (2020)},'' \emph{DOI}, vol.~10, p. 77482,
  2020.

\bibitem{RasSchBarVilc20}
R.~Raskar, I.~Schunemann, R.~Barbar, K.~Vilcans, J.~Gray, P.~Vepakomma,
  S.~Kapa, A.~Nuzzo, R.~Gupta, A.~Berke \emph{et~al.}, ``{Apps Gone Rogue:
  Maintaining Personal Privacy in an Epidemic},'' \emph{arXiv preprint
  arXiv:2003.08567}, 2020.

\bibitem{Quarantine20}
\BIBentryALTinterwordspacing
``{COVID-19 und h\"ausliche Quarant\"ane: Flyer f\"ur Gesundheits\"amter},''
  Robert Koch Institut, Berlin, Tech. Rep., 2020. [Online]. Available:
  \url{https://www.rki.de/DE/Content/InfAZ/N/Neuartiges_Coronavirus/Quarantaene/Flyer.pdf?__blob=publicationFile}
\BIBentrySTDinterwordspacing

\bibitem{MoaLiShi19}
N.~Moayeri, C.~Li, and L.~Shi, ``Indoor localization accuracy of major
  smartphone location apps,'' in \emph{2019 IEEE Wireless Communications and
  Networking Conference (WCNC)}.\hskip 1em plus 0.5em minus 0.4em\relax IEEE,
  2019, pp. 1--8.

\bibitem{Category20}
\BIBentryALTinterwordspacing
``{Kontaktpersonennachverfolgung bei respiratorischen Erkrankungen durch das
  Coronavirus SARS-CoV-2},'' Robert Koch Institut, Berlin, Tech. Rep., 2020.
  [Online]. Available:
  \url{https://www.rki.de/DE/Content/InfAZ/N/Neuartiges_Coronavirus/Kontaktperson/Management_Download.pdf?__blob=publicationFile}
\BIBentrySTDinterwordspacing

\end{thebibliography}

\end{document}